\documentclass[10pt]{revtex4}
\newcount\driver
\newcount\bozza

\font\ottorm=cmr8\font\ottoi=cmmi8\font\ottosy=cmsy8%
\font\ottocss=cmcsc8%
\font\sixrm=cmr6\font\sixi=cmmi6\font\sixsy=cmsy6%
\font\fiverm=cmr5\font\fivesy=cmsy5
\font\fivei=cmmi5
\font\tenmib=cmmib10
\font\sevenmib=cmmib10 scaled 800

 2

\font\sc=cmcsc10

\font\ss=cmss10
\font\sss=cmss8

\font\elevenrm=cmr11
\font\twelverm=cmr12
\font\ottorm=cmr8
\font\msytw=msbm9 scaled\magstep1

\font\msytwww=msbm5 scaled\magstep1
\font\indbf=cmbx10 scaled\magstep2

\font\ottorm=cmr8\font\ottoi=cmmi8\font\ottosy=cmsy8%
\font\ottocss=cmcsc8%
\font\sixrm=cmr6\font\sixi=cmmi6\font\sixsy=cmsy6%
\font\fiverm=cmr5\font\fivesy=cmsy5
\font\fivei=cmmi5

\def\ottopunti{\def\rm{\fam0\ottorm}%
\textfont0=\ottorm\scriptfont0=\sixrm\scriptscriptfont0=\fiverm%
\textfont1=\ottoi\scriptfont1=\sixi\scriptscriptfont1=\fivei%
\textfont2=\ottosy\scriptfont2=\sixsy\scriptscriptfont2=\fivesy%
\textfont4=\ottocss\scriptfont4=\sc\scriptscriptfont4=\sc%
\scriptfont4=\ottocss\scriptscriptfont4=\ottocss%
\textfont5=\tenmib\scriptfont5=\sevenmib\scriptscriptfont5=\fivei
\setbox\strutbox=\hbox{\vrule height7pt depth2pt width0pt}%
\normalbaselineskip=9pt\let\sc=\sixrm\normalbaselines\rm}

\mathchardef\BDpr = "0540  
\mathchardef\Bg   = "050D  

{\count255=\time\divide\count255 by 60 \xdef\hourmin{\number\count255}
        \multiply\count255 by-60\advance\count255 by\time
   \xdef\hourmin{\hourmin:\ifnum\count255<10 0\fi\the\count255}}

\def\openone{\leavevmode\hbox{\elevenrm 1\kern-3.63pt\twelverm1}}%
\def\*{\vglue0.5truecm}

\let\a=\alpha \let\b=\beta  \let\g=\gamma  \let\d=\delta \let\e=\varepsilon
     \let\th=\theta  
    \let\n=\nu             
\let\s=\sigma \let\t=\tau    
   
\let\G=\Gamma \let\D=\Delta  \let\L=\Lambda

\def\\{\hfill\break} \let\==\equiv

\let\io=\infty 

\let\0=\noindent

\def\ie{{i.e. }}\def\eg{{e.g. }}
\let\dpr=\partial

\def\tende#1{\,\vtop{\ialign{##\crcr\rightarrowfill\crcr
 \noalign{\kern-1pt\nointerlineskip} \hskip3.pt${\scriptstyle
 #1}$\hskip3.pt\crcr}}\,}
\def\circage{\lower2pt\hbox{$\,\buildrel > \over {\scriptstyle \sim}\,$}}
\def\otto{\,{\kern-1.truept\leftarrow\kern-5.truept\to\kern-1.truept}\,}
\def\fra#1#2{{#1\over#2}}

 \def\III{{\cal I}}
  
\def\DD{{\cal D}}

\def\T#1{{#1_{\kern-3pt\lower7pt\hbox{$\widetilde{}$}}\kern3pt}}
\def\VVV#1{{\underline #1}_{\kern-3pt
\lower7pt\hbox{$\widetilde{}$}}\kern3pt\,}
\def\W#1{#1_{\kern-3pt\lower7.5pt\hbox{$\widetilde{}$}}\kern2pt\,}

\def\indica{\leaders \hbox to 0.5cm{\hss.\hss}\hfill}
\def\guida{\leaders\hbox to 1em{\hss.\hss}\hfill}

\def\hhh{{\bf h}}

\def\ul{\underline}

\def\defin{{\buildrel def\over=}}
\def\wt{\widetilde}

\mathchardef\aa   = "050B
\mathchardef\bb   = "050C
\mathchardef\ggg  = "050D
\mathchardef\xxx  = "0518
\mathchardef\zzzzz= "0510
\mathchardef\oo   = "0521
\mathchardef\lll  = "0515
\mathchardef\mm   = "0516
\mathchardef\Dp   = "0540
\mathchardef\H    = "0548
\mathchardef\FFF  = "0546
\mathchardef\ppp  = "0570
\mathchardef\Bn   = "0517
\mathchardef\pps  = "0520
\mathchardef\fff  = "0527
\mathchardef\FFF  = "0508
\mathchardef\nnnnn= "056E

\def\to{\rightarrow}

\def\qed{\hfill\raise1pt\hbox{\vrule height5pt width5pt depth0pt}}

\def\indic{\hbox{\raise-2pt \hbox{\indbf 1}}}

\def\RRR{\hbox{\msytw R}}

\def\nnn{\hbox{\msytwww N}} \def\ZZZ{\hbox{\msytw Z}}
 \def\zzz{\hbox{\msytwww Z}}
\def\SSS{\hbox{\msytw S}}
 \def\sss{\hbox{\msytwww S}}

\def\ul#1{{\underline#1}}

\def\V0{{\bf 0}}

\font\tenmib=cmmib10 
\font\sevenmib=cmmib7\font\fivemib=cmmib5 

\font\fivei=cmmi5\font\sixi=cmmi6\font\ottoi=cmmi8
\font\ottorm=cmr8\font\fiverm=cmr5\font\sixrm=cmr6
\font\ottosy=cmsy8\font\sixsy=cmsy6\font\fivesy=cmsy5
\font\ottocss=cmcsc8%

\mathchardef\Ba   = "050B  
\mathchardef\Bb   = "050C  
\mathchardef\Bg   = "050D  
\mathchardef\Bd   = "050E  
\mathchardef\Be   = "0522  
\mathchardef\Bee  = "050F  
\mathchardef\Bz   = "0510  
\mathchardef\Bh   = "0511  
\mathchardef\Bthh = "0512  
\mathchardef\Bth  = "0523  
\mathchardef\Bi   = "0513  
\mathchardef\Bk   = "0514  
\mathchardef\Bl   = "0515  
\mathchardef\Bm   = "0516  
\mathchardef\Bn   = "0517  
\mathchardef\Bx   = "0518  
\mathchardef\Bom  = "0530  
\mathchardef\Bp   = "0519  
\mathchardef\Br   = "0525  
\mathchardef\Bro  = "051A  
\mathchardef\Bs   = "051B  
\mathchardef\Bsi  = "0526  
\mathchardef\Bt   = "051C  
\mathchardef\Bu   = "051D  
\mathchardef\Bf   = "0527  
\mathchardef\Bff  = "051E  
\mathchardef\Bch  = "051F  
\mathchardef\Bps  = "0520  
\mathchardef\Bo   = "0521  
\mathchardef\Bome = "0524  
\mathchardef\BG   = "0500  
\mathchardef\BD   = "0501  
\mathchardef\BTh  = "0502  
\mathchardef\BL   = "0503  
\mathchardef\BX   = "0504  
\mathchardef\BP   = "0505  
\mathchardef\BS   = "0506  
\mathchardef\BU   = "0507  
\mathchardef\BF   = "0508  
\mathchardef\BPs  = "0509  
\mathchardef\BO   = "050A  
\mathchardef\BDpr = "0540  
\mathchardef\Bstl = "053F  

\def\V#1{{\bf#1}}
\let\aa=\Ba\let\fff=\Bf\let\defin=\defi
\let\wt=\widetilde\let\oo=\Bo\let\nn=\Bn
\let\pps=\Bps\def\hhh={\V h}
\let\bb=\Bb\def\ss{\ul{\s}}

\def\RRR{\hbox{\msytw R}}

\def\nnn{\hbox{\msytwww N}} \def\ZZZ{\hbox{\msytw Z}}
 \def\zzz{\hbox{\msytwww Z}}

\let\ul=\underline

\def\ins#1#2#3{\vbox to0pt{\kern-#2 \hbox{\kern#1 #3}\vss}\nointerlineskip}
\newdimen\xshift \newdimen\xwidth \newdimen\yshift
\newcount\griglia

\def\insertplot#1#2#3#4#5#6{%
\begin{figure}[h]
\begin{center}
\vspace{#2pt}
\begin{minipage}{#1pt}
#3
\ifnum\driver=1
\griglia=#6
\ifnum\griglia=1
\openout13=griglia.ps
\write13{gsave .2 setlinewidth}
\write13{0 10 #1 {dup 0 moveto #2 lineto } for}
\write13{0 10 #2 {dup 0 exch moveto #1 exch lineto } for}
\write13{stroke}
\write13{.5 setlinewidth}
\write13{0 50 #1 {dup 0 moveto #2 lineto } for}
\write13{0 50 #2 {dup 0 exch moveto #1 exch lineto } for}
\write13{stroke grestore}
\closeout13
\includegraphics{griglia.ps}\fi
\includegraphics{#4.ps}\fi
\ifnum\driver=2
\fi
\end{minipage}
\end{center}
\caption{#5}
\end{figure}
}

\newdimen\shift \shift=-1truecm
\def\lb#1{%
\ifnum\bozza=1
\label{#1}\rlap{\kern\shift{$\scriptstyle#1$}}
\else\label{#1}
\fi}

\def\be{\begin{equation}}
\def\ee{\end{equation}}
\def\bea{\begin{eqnarray}}\def\eea{\end{eqnarray}}
\def\bean{\begin{eqnarray*}}\def\eean{\end{eqnarray*}}
\def\bfr{\begin{flushright}}\def\efr{\end{flushright}}
\def\bc{\begin{center}}\def\ec{\end{center}}
\def\ba#1{\begin{array}{#1}} \def\ea{\end{array}}
\def\bd{\begin{description}}\def\ed{\end{description}}
\def\bv{\begin{verbatim}}\def\ev{\end{verbatim}}
\def\nn{\nonumber}
\def\Halmos{\hfill\vrule height10pt width4pt depth2pt \par\hbox to \hsize{}}

\renewcommand{\theequation}{\arabic{section}.\arabic{equation}}

\newdimen\xshift \newdimen\xwidth \newdimen\yshift \newdimen\ywidth

\def\ins#1#2#3{\vbox to0pt{\kern-#2\hbox{\kern#1 #3}\vss}\nointerlineskip}

\def\eqfig#1#2#3#4#5{
\par\xwidth=#1 \xshift=\hsize \advance\xshift
by-\xwidth \divide\xshift by 2
\yshift=#2 \divide\yshift by 2
\line{\hglue\xshift \vbox to #2{\vfil
#3 \includegraphics{#4.ps}
}\hfill\raise\yshift\hbox{#5}}}

\def\8{\write12}

\begin{document}

\title{Ising models with long-range antiferromagnetic\\ and short-range
ferromagnetic interactions}
\*

\author{Alessandro Giuliani}
\affiliation{Department of Physics, 
Princeton University, Princeton 08544 NJ, USA}
\author{Joel L. Lebowitz}
\affiliation{Department of Mathematics and Physics, Rutgers University,
Piscataway, NJ 08854 USA.}
\author{Elliott H. Lieb}
\affiliation{Departments of Mathematics and Physics, 
Princeton University, Princeton, NJ 08544 USA.}
\vspace{1cm}
\date{\today}
\begin{abstract} 
We study the ground state of a $d$--dimensional Ising model with both long 
range (dipole--like) and nearest neighbor ferromagnetic (FM) interactions. 
The long range interaction is equal to $r^{-p}$, $p>d$, while the FM 
interaction has strength $J$. If $p>d+1$ and $J$ is large enough the ground 
state is FM, while if $d<p\le d+1$ the FM state is not the ground state for 
any choice of $J$. In $d=1$ we show that for any $p>1$ the ground state has a 
series of transitions from an antiferromagnetic state of period 2 
to $2h$--periodic states of blocks of sizes $h$ with alternating sign, 
the size $h$ growing when the FM interaction strength $J$ is increased
(a generalization of this result to the case $0<p\le 1$ is also discussed).
In $d\ge 2$ we prove, for $d<p\le d+1$, that the dominant asymptotic 
behavior of the ground state energy agrees for large $J$ with that obtained
from a periodic striped state conjectured to be the true ground state.
The geometry of contours in the ground state is discussed.
\end{abstract}

\maketitle

\section{Introduction}

There has been and continues to be much interest
in mesoscopic pattern formation induced by competing short and long range 
interactions \cite{B01,M02,BN98,MP03}. 
Of particular interest is the competition between 
dipole--like interactions bewteen spins which decay as $r^{-(d+1)}$, $d$ the 
dimension of the lattice, and short range exchange interactions 
\cite{SA95,SS99}. For $d=2$, 
this type of competitive interaction is believed responsible for many of the 
observed patterns in thin magnetic films \cite{SW92} and other 
quasi--two dimensional systems, including Langmuir monolayers \cite{M88},
lipid monolayers \cite{KM99}, liquid crystals \cite{MS92}, polymer films
\cite{H00} and two--dimensional electron gases \cite{SK04,SK06}. 

The simplest models to describe such systems are Ising spins with a
nearest neighbor ferromagnetic interaction and a power law long range 
antiferromagnetic pair potential, of dipole (or Coulomb) type
\cite{MWRD95,AWMD95,SS99,LEFK94,GTV00,GD82,CEKNT96}. 
The zero temperature phase diagram of these models has been thoroughly
investigated over the last decade and a sequence of transitions 
from an antiferromagnetic N\'eel state to periodic striped or lamellar phases 
with domains of increasing sizes has been predicted, as 
the strength of the ferromagnetic coupling is increased from zero to large
positive values. These theoretical predictions are mostly based on 
a combination of variational techniques and stability analysis: they start by
{\it assuming} a periodic structure, proceed by computing the 
corresponding energy and finally by comparing that energy to the 
energy of other candidate structures, usually by a combination of analytical
and numerical tools. These calculations give an 
excellent account of the observed ``universal'' patterns displayed by 
the aforementioned quasi--two dimensional systems. However they run the risk 
of overlooking complex microphases that have not been previously identified 
\cite{BF99}. 
This risk is particularly significant in cases, as those under analysis,
where dynamically (e.g. in Monte Carlo simulations)
the domain walls separating different microphases appear to 
be very long lived as the temperature is lowered \cite{MWRD95}.

In order to settle the question of spontaneous pattern formations
on more solid grounds it would be desirable to be able to first {\it prove} 
periodicity of the ground state and then proceed with 
a variational computation within the given ansatz. The problem is not simple. 
Most 
of the mathematically rigorous techniques developed for obtaining the low 
temperature phase diagram of spin systems, \eg the Pirogov--Sinai theory 
\cite{PS75}, depend on the interaction being short range. Only methods based 
on reflection positivity \cite{FILS78} or on convexity \cite{Hu78,PU78,JM001,
JM002,K99} seem applicable to the kind of systems considered here.
In this paper we apply reflection positivity to give a characterization of 
the ground states of the one--dimensional system and to give in this 
case a full justification of the variational calculation based on the 
periodicity assumption. Moreover we obtain 
rigorous upper and lower bounds on the ground state energy in higher 
dimensions, thus providing a quantitative estimate of the possible 
metastability of the striped or lamellar phases. 

\subsection{The model}

Given an integer $N$, let $\L_N$ be a ``simple cubic'' 
$d$--dimensional torus of side $2N$
and $\ss\in\{\pm 1\}^{\L_N}$. Any configuration $\ss$ is a sequence
of $\s_i=\pm 1$ labelled by $i\in \L_N$.
The Hamiltonian of this Ising model with {\it periodic boundary conditions} 
is taken to be 
\begin{equation}H_N(\ss)=-J\sum_{i\in\L_N}\sum_{k=1}^d\s_i\s_{i+e_k} 
+\sum_{(i,j)} 
\s_i\, J_p(j-i)\; \s_j\;,\qquad J_p(j-i)=\sum_{n\in \zzz^d}
\fra1{|i-j+2n N|^p}\label{1.1}\end{equation}
where in the first sum $e_k$ is the unit vector in the $k$--th 
coordinate direction (and if $i+e_k\not\in\L_N$ we define $\s_{i+e_k}\=
\s_{i-2Ne_k}$), the second sum runs over generic distinct pairs of sites 
in $\L_N$ and $p>d$.
The first term in 
(\ref{1.1}) will be called the {\it exchange energy} term and we will take 
$J\ge 0$ while the second one 
will be called the {\it dipolar energy} term. In $d>1$ we shall define 
$|i-j|$ to be the usual Euclidean distance between $i$ and $j$.
Note that the sum over $n$ in the definition of $J_p(j-i)$ makes sense 
as long as $p>d$. However some of the results discussed below hold true even 
in the case $d-1< p\le d$, {\it provided} the definition of $J_p(j-i)$
is replaced by $J_p(j-i)=|j-i|^{-p}+(2N-|j-i|)^{-p}$, $\forall 1\le |j-i|\le 
2N-1$. In the following we will restrict attention to the case $p>d$
and we will comment on possible generalizations of our results to $d-1<p\le d$ 
in the Remark after Theorem 2. See also the Remark after Theorem 3.

\subsection{Main results}

We wish to determine the ground state of model
(\ref{1.1}) for different values of $p$ and $J$. 
A first remark is that in any dimension for $p>d+1$ and $J$ large enough
the ground state is ferromagnetic and is unique modulo a global spin flip.
This is, in $d>1$, a corollary of the contour estimates in \cite{GGR66}, and 
we shall reproduce its proof as a byproduct
of our analysis. For this case the low 
temperature Gibbs states are also known: for $d=1$ and $p>2$ there is
a unique Gibbs state for any $\b<+\io$ \cite{R68} 
while for $d\ge 2$, $p>d+1$ and
$\b$ large enough, there are two different pure states obtained as the 
thermodynamic limits of the equilibrium 
states with $+$ or $-$ boundary conditions 
\cite{GGR66}.
On the contrary, for $d<p\le d+1$, the ferromagnetic state
is not the ground state in any dimension, for any value of $J$. 
Instability of the ferromagnetic state in presence of a long range 
antiferromagnetic interaction of the form $1/r^p$, $d<p\le d+1$,
was first noted in \cite{VE81}. The reason for this instability 
lies in the divergence of the first moment of the long range interaction 
which makes the ferromagnetic state unstable towards flipping spins in a large 
enough domain.

Our main results give a characterization of the ground state in the whole 
regime $p>d$. For $d=1$ the characterization is complete, in the sense that 
for any $p>1$ we can compute the ground state energy per site and
we can prove that the corresponding 
ground state configurations are periodic and consist 
of blocks of equal size and alternating signs 
(a block is a maximal sequence of adjacent spins of the same sign).
\\
\\
{\bf Theorem 1.} (Ground state energy, d=1).\\
{\it For $d=1$ and $p>1$, 
the ground state energy $E_0(N)$ of the Hamiltonian (\ref{1.1}) 
satisfies
\begin{equation}\lim_{N\to\io}\frac1{2N} E_0(N)=\min_{h\in\zzz^+}e(h)
\label{1.2a}\end{equation}
where $e(h)\defin\lim_{N\to\io} (2Nh)^{-1}E_{per}^{(h)}(Nh)$  and
$E_{per}^{(h)}(Nh)$ is the energy of a periodic configuration
on a ring of $2Nh$ sites consisting of blocks of size $h$ with
alternating signs.}
\\

The function $e(h)$ is explicitly computable, see Section 
\ref{sec3}. The minimization of $e(h)$ with respect to $h$ can be 
performed exactly and, as expected, for $p>2$ and 
$J$ larger than an explicit constant 
$J_0\=\G(p)^{-1}\int_0^\io d\a\a^{p-1}e^{-\a}(1-e^{-\a})^{-2}$,
the minimizer is $h=+\io$ (\ie the ground state is
ferromagnetic). For $1<p\le 2$ and all $J\ge 0$ or $p>2$ and $0\le J<J_0$
the minimum of $e(h)$ over the positive integers is attained at
an integer $h^*(J)$ which is a piecewise constant monotone increasing 
function of $J$. Thus 
there is a unique integer minimizer of $e(h)$ for almost all values of $J$
which jumps at a discrete set of values of $J$. At those values of $J$ the 
minimum is attained at two consecutive integers $h^*(J)$ and $h^*(J)+1$
and the corresponding ground states, for suitable values of $N$,
are the configurations with all blocks of the same
size, either $h^*(J)$ or $h^*(J)+1$.
\\
\\
{\bf Theorem 2.} (Finite volume ground states).\\
{\it Let $d=1$ and $p>1$. At the values of $J$ such that 
the minimum of $e(h)$ is attained at a single integer $h^*(J)$,
then in a ring of length $2N$, such that $N$ is divisible by 
$h^*(J)$, the only ground states are the periodic configurations
consisting of blocks of size $h^*(J)$. At the values of $J$ such that the 
minimum is attained at two consecutive integers $h^*(J)$ and $h^*(J)+1$,
then in a ring of length $2N$, such that $N$ is divisible both by $h^*(J)$
and $h^*(J)+1$, the only ground states are the periodic configurations 
consisting of blocks of the same size, either $h^*(J)$ or $h^*(J)+1$.
In both cases there is a finite (independent of $N$) gap between the 
energies of the ground states and of any other state.}\\
\\
{\it Remarks.}\\ 
\01) {\it Stability for $0<p\le 1$}. The proofs of Theorems 1 and 2
do not really require $p>1$ and, provided we change the definition of 
$J_p(j-i)$ as described after (\ref{1.1}), they can be easily adapted to
cover the cases $0<p\le 1$. In particular for any $p>0$ 
the ground state energy can be 
computed by minimizing energy among the periodic
states of blocks with alternating signs and the computation shows that
the ground state energy is {\it extensive} 
(\ie it scales proportionally to $N$ as $N\to \io$), even for 
$0<p\le 1$ (a case in which thermodynamic stability is not a priori 
guaranteed).\\
\02) {\it Infinite volume ground states}.
A corollary of Theorem 2 is that the infinite volume periodic 
configurations $\ss_J\in\{\pm1\}^{\zzz}$ 
with blocks all of the same size $h^*(J)$ (or $h^*(J)+1$,
if $J$ is a value at which the minimum of $e(h)$ is not unique)
and alternating sign 
are infinite volume ground state configurations, in the sense
that they are stable against bounded perturbations \cite{PS75}; 
\ie given a finite set $X\subset\ZZZ$, 
if $R_X$ is the operator flipping the spins in
$X$, the (finite) energy difference between $R_X\ss_J$ and $\ss_J$ 
is {\it positive}. This can be proven by writing the energy difference
between the infinite configurations $R_X\ss_J$ and $\ss_J$ as the limit of the
energy differences between the finite volume configurations on rings
of length $2N$ ($N$ divisible by $h^*(J)$) obtained by restricting 
$R_X\ss_J$ and $\ss_J$ to the finite rings. By Theorem 2, the differences 
between the energies of the finite volume approximations of $R_X\ss_J$
and $\ss_J$ are positive. This implies that $\ss_J$ is an infinite volume 
ground state configuration. 

A stronger result which follows from our analysis is that for 
any $J\ge 0$ we can choose a sequence 
of integers $\{N_i\}_{i\in\nnn}$, $N_i\tende{i\to\io}\io$, 
such that the ground states on the rings of lenghts 
$2N_i$ are periodic 
configurations of blocks of size $h^*(J)$ and the corresponding sequence of
ground state energies per site $e_0(N_i)$ is the sequence of ``lowest
possible specific energies'', \ie any other sequence
$\{N_i'\}_{i\in\nnn}$, $N_i'\tende{i\to\io}\io$, has $e_0(N_i')\ge e_0(N_i)$, 
eventually as $i\to\io$.
\\
\\
In dimension larger then one we do not find an exact asymptotic 
expression for the ground state energy. Still, for large $J$, we
find rigorous upper and lower bounds to the ground state energy. The result 
is the following.\\
\\
{\bf Theorem 3.} {\it Let $d\ge2$.
If $p=d+1$ and $J$ is large enough,  
there exist positive, $J$--independent, constants $C_1,C_2$ such that 
\bea&&
e_{FM}(J)-C_1 e^{-J|\sss_{d-1}|^{-1}}\le\lim_{N\to\io}\frac1{|\L_N|} 
E_0(N)\le e_{FM}(J)-C_2 e^{-J|\sss_{d-1}|^{-1}}\;,\label{1.3aa}\eea
where $e_{FM}(J)$ is the energy per site of the ferromagnetic 
configuration $(\ss)_i=+1$ and $|\SSS_{d-1}|$ is the volume of the 
$d-1$ dimensional unit sphere. If $d<p<d+1$ and $J$ is large enough, then 
there exist positive, $J$--independent, constants $K_1,K_2$ such that
\bea&&e_{FM}(J)-K_1 J^{-\frac{p-d}{d+1-p}}\le\lim_{N\to\io}\frac1{|\L_N|} 
E_0(N)\le e_{FM}(J)-K_2 J^{-\frac{p-d}{d+1-p}}\;.\label{1.3}\eea}

\\
{\it Remark.} ({\it The case $p\le d$}). 
Contrary to the proofs of Theorem 1 and 2, 
the proof of Theorem 3 crucially relies on the summability of the potential,
\ie on the condition $d<p\le d+1$. It would be of great interest to extend the 
proof to the case $d-1<p\le d$ and to the case $p=1$ in dimension
$d=2,3$ (Coulomb case). 
A thorough discussion of the case $d-1\le p\le d+1$ in dimension
$d\ge 2$, including results related to those of Theorem 3,
is provided by Spivak and Kivelson \cite{SK06}. 
The special case $p=1$ with $d=2$ was treated 
in \cite{LEFK94}, and in full generality in \cite{JKS05}. The Coulomb case  
$p=1$ in $d=3$ was considered in \cite{GTV00}. 
\\
\\
The upper bounds in (\ref{1.3aa}) and (\ref{1.3}) follow from a 
variational computation. The best known constants $C_2,K_2$ are obtained 
by minimizing over the periodic striped configurations
$(\ss_{striped}^{(h)})_{i}=(-1)^{[i_1/h]}$, where $i_1$ is the first 
component of $i$ and $[x]$ is the largest integer less than or equal to $x$.
It is remarkable that, for $J\to\io$, the minimum over the 
striped configurations provides a variational energy that is lower than the 
one obtained by minimizing over the periodic checkerboard configurations;
this was shown in \cite{MWRD95} for $d=2$ and $p=3$. A computation 
analogous to that in \cite{MWRD95} allows to prove a similar result in higher 
dimensions for $d<p\le d+1$.
This result together with numerical studies support the conjecture 
that the ground state of (\ref{1.1}) in $d\ge2$ and $d<p\le d+1$ is in fact a 
periodic striped configuration (at least asymptotically for large $J$).
In this paper we prove the lower bounds in (\ref{1.3aa}) and (\ref{1.3}), 
which show that for large $J$ the difference between the 
ground state energy and the ferromagnetic energy scales exactly as predicted 
by the above mentioned variational computation. The proof of the lower bound 
is based on an energy argument which also gives estimates on the 
allowed shapes and sizes of Peierls' contours in the ground state.

\*\section{Outline of the proofs}

Before presenting the proofs of Theorems 1, 2 and 3 in detail, we give a 
short outline. 

The proofs of Theorem 1 and 2 are based on {\it reflection positivity}.
It is known \cite{FILS80} that for $J=0$ the Hamiltonian 
(\ref{1.1}) is reflection positive in any dimension. In $d=1$ this means that, 
given any configuration of spins on a ring of $2N$ sites
$\ss=(\s_{-N+1},\ldots,\s_{-1},\s_0,\s_1,\s_2,\ldots,\s_N)$ and a bond 
$b=(i,i+1)$ connecting site $i$ with its neighbor $i+1$, 
the average energy of the two configurations obtained by reflecting around 
$b$ is always lower than the original one; \eg choosing $b=(0,1)$ 
the average of the energies of the two reflected configurations
$(-\s_{N}\ldots,-\s_{2},-\s_1,\s_1,\s_2,\ldots,\s_N)$ and
$(\s_{-N+1}\ldots,\s_{-1},\s_0,-\s_0,-\s_{-1},\ldots,-\s_{-N+1})$
is always lower than the energy of the original configuration $\ss$. 
Using this property
and repeatedly reflecting around different bonds, one can prove that 
for $J=0$ the ground state is the antiferromagnetic alternating state.
Note that periodic boundary conditions are necessary for repeated reflections 
around different bonds.

Unfortunately, as soon as $J>0$, the Hamiltonian (\ref{1.1}) is {\it not}
reflection positive anymore. Still, one can think of making 
use of the reflection symmetry around bonds $b=(i,i+1)$ separating a spin 
$\s_i$ from a spin $\s_{i+1}=-\s_i$: such reflection does not change the 
exchange energy between $\s_i$ and $\s_{i+1}$ and lowers the dipole energy.
By repeatedly reflecting around 
such sites one could expect to be able to reduce the search for the ground 
states just to the class of periodic configurations of blocks of the same 
size and alternating sign and explicitly look for periodic configuration 
with minimal energy (we recall that by block we mean a maximal 
sequence of consecutive spins all of the same sign).
However there is a difficulty: because of periodic boundary conditions,
in order not to increase the exchange energy in the reflection, 
one should reflect around bonds $b$ not only separating a $+$ from a $-$ 
spin, but with the further 
property that the bond $b'$ at a distance $N$ from $b$
also separates a $+$ from a $-$ spin: but in a generic configuration $\ss$ 
there will be no bond $b$ with such a property! 

A possible solution to this problem is to cut the ring into two uneven 
parts both containing the same number of blocks but not necessarily 
of the same length. The configurations obtained by reflecting the two 
uneven parts will have a lower energy but in general different lengths 
$2N',2N''$. The idea is to reflect repeatedly in this fashion, keeping 
track of the errors due to the fact that the length of the ring is
changing at each reflection. A convenient way of doing this, exploited in 
Section \ref{sec3}, 
is to rewrite the spin Hamiltonian (\ref{1.1}) as an effective 
Hamiltonian for new ``atoms'' of ``charges'' $h_i$, corresponding to the 
spin blocks of size $h_i$. The new effective Hamiltonian 
$E(\ldots,h_{-1},h_0,h_1,h_2,\ldots)$ is again reflection symmetric 
(in a slightly different sense, though) and 
its explicit form allows for an easy control of the finite size errors
one is left with after repeated reflections. Some technical aspects of the 
proofs of Theorem 1 and 2 are given in the Appendices.

The proof of Theorem 3, in particular the lower bounds in (\ref{1.3aa})
and (\ref{1.3}), is based on a Peierls' contour estimate described in Section 
\ref{sec4}. For large ferromagnetic coupling $J$,
we shall describe the ground state configuration(s) in terms of droplets 
of $-$ spins surrounded by $+$ spins and of contours separating the $+$ 
from the $-$ phases. 
The requirement that the energy of the ground state configuration is minimal 
imposes bounds on the geometry and the energy of such droplets,
implying the lower bound in (\ref{1.3}). Contrary to the methods exploited 
in the proofs of Theorem 1 and 2, the proof of Theorem 3 is robust under 
modifications both of the boundary conditions and of the specific form
of the interaction potential.

\*\section{One dimension}\lb{sec3}

In this section we want to prove Theorems 1 and 2.
Restricting to $d=1$, 
we shall consider any configuration $\ss\in\{\pm1\}^{\L_N}$ on the ring
of length $2N$
as a sequence of blocks of alternating sign, where by definition a block is
a maximal sequence of adjacent spins of the same sign.
We note that, due to the periodic boundary condition, the number $M$ of blocks
on the ring is either $1$ (if the state is ferromagnetic) or an even number. 
We shall denote by $h_i$, $i=1,\ldots,M$, the sizes of such blocks.
A block will be called a $+$ block ($-$ block), if its spins are all of sign 
$+1$ ($-1$).

We want to prove that the sizes $h_i$ of the blocks in the ground 
state are all equal, at least for $N$ large enough.
The strategy will be the following. 
Given any configuration $\ss_N$ in the ring $\L_N$ with $M\ge 2$ blocks of 
alternating signs of sizes $h_{-\fra{M}2+1},\ldots,h_{\fra{M}2}$,
we will rewrite the energy $H_N(\ss_N)$ as a function of $M$ and the $h_i$'s,
\ie $H_N(\ss_N)=E_N\big(M,(\ul h_L,\ul h_R)\big)$,
where $\ul h_L=(h_{-\fra{M}2+1},\ldots,h_0)$ and $\ul h_R=(h_1,\ldots,
h_{\fra{M}2})$. Setting $\ul h=(\ul h_L,\ul h_R)$, we will show that, 
for $M$ and $N$ fixed, the Hamiltonian $E_N(M,\ul h)$ is 
{\it reflection positive} with respect to the reflection 
\begin{equation}\hat\th h_i=h_{-i+1}\;,
\qquad -\fra{M}2<i\le \fra{M}2\;,\label{3.12}\end{equation}
that is 
\begin{equation}E_N\big(M,\ul h\big)\ge \frac12
\left[E_N \big(M,(\ul h_L,\hat\th\ul h_L)\big)+E_N
\big(M,(\hat\th\ul h_R,\ul h_R)\big)\right]\;,\label{3.12a}\end{equation}
where $\hat\th\ul h_L=(h_0,\ldots,h_{-\fra{M}2+1})$ and 
$\hat\th\ul h_R=(h_{\fra{M}2},\ldots,h_1)$, with of course 
the signs of the spins in the reflected blocks being opposite to what they were
originally.
Using repeatedly this symmetry, we will get bounds from above and below
for the ground state energy, in terms of the energy of periodic configurations.
\subsection{The integral representation}
In order to show reflection positivity of $E_N(M,\ul h)$, \eg (\ref{3.12a}),  
we need to look more closely at the structure of the Hamiltonian.
A straightforward calculation in Appendix \ref{A4}
gives the energy $E_N(M,\ul h)$ of a configuration in $\L_N$ of $M$ blocks of 
sizes $h_{-\fra{M}{2}+1},\ldots,h_{\fra{M}{2}}$ (recall that 
$M$ is even and p.b.c. are assumed) as:
%
\bea&&
E_N(M,\ul h)=-2J(N-M)+2N\int_0^\io 
\fra{d\a}{\G(p)}\,\a^{p-1}\fra{e^{-\a}}{(1-e^{-\a})^2(1-e^{-2\a N})}
\cdot\nn\\
&&\cdot\Big\{
\sum_{i=-\fra{M}{2}+1}^{\fra{M}{2}}\big[h_i(1-e^{-\a})(1-e^{-2\a N})-
(1-e^{-\a h_i})(1-2 e^{-\a(2N-h_i)}+
e^{-2\a N})\big]+\nn\\
&&+\sum_{-\fra{M}{2}< i<j\le \fra{M}{2}}(-1)^{j-i}(1-e^{-\a h_i})
(1-e^{-\a h_j})\big[\prod_{i<k<j}e^{-\a h_k}+\prod_{-\fra{M}{2}< k<i\atop
j<k\le \fra{M}{2}}e^{-\a h_k}\big]\Big\}\;.\label{3.11}\eea

The expression in braces in the last two lines of (\ref{3.11}) can be 
pictorially interpreted as the energy of a system of particles 
on a ring of $M$ sites with $h_i\ge 1$ particles at site $i$. 
There is an on site energy as in the second line of (\ref{3.11}) and a 
``many body'' energy as in the third line. 

The energy in (\ref{3.11}) has the 
remarkable property of being {\it reflection positive} with respect to
the reflection (\ref{3.12}). This can be proven by rewriting the 
expression in braces in the last two lines of (\ref{3.11}) as
\begin{equation}\int_0^\io
\n(\a)\,\Big[H_R(\a,\ul h_R)+H_R(\a,\hat\th\ul h_L)-\sum_{i=1}^2
F_i(\a, \ul h_R)\,F_i(\a,\hat\th \ul h_L)\Big]d\a\;,\label{3.13}\end{equation}
where
\bea&& \n(\a)\=\fra{1}{\G(p)}
\,\a^{p-1}\fra{e^{-\a}}
{(1-e^{-\a})^2(1-e^{-2\a N})}\;,\nn\\
&&H_R(\a,\ul h_R)=-\sum_{i=1}^{\fra{M}{2}}
(1-e^{-\a h_i})(1-2 e^{-\a(2N-h_i)}+e^{-2\a N})+\nn\\
&&\hskip1.8truecm+\sum_{1\le i<j\le \fra{M}{2}}(-1)^{j-i}(1-e^{-\a h_i})
(1-e^{-\a h_j})\prod_{i<k<j}e^{-\a h_k}
\;,\nn\\
&& 
F_1(\a, \ul h_R)=\sum_{i=1}^{\fra{M}{2}}(-1)^i(1-e^{-\a h_i})\prod_{1\le k<i}
e^{-\a h_k}\;,\nn\\
&&F_2(\a, \ul h_R)=\sum_{i=1}^{\fra{M}{2}}(-1)^i(1-e^{-\a h_i})\prod_{i<k\le 
\fra{M}2}e^{-\a h_k}\;,\label{3.14}\eea
Let us define $H(\a,\ul h)$ as follows:
\be H(\a,\ul h)=H_R(\a,\ul h_R)+H_R(\a,\hat\th \ul h_L)-\sum_{i=1}^2
F_i(\a, \ul h_R)\, F_i(\a,\hat\th \ul h_L)\label{3.14aa}\ee
so that $E_N(M,\ul h)=-2J(N-M)+2N\int_0^\io 
\fra{d\a}{\G(p)}\,\a^{p-1}\fra{e^{-\a}}{(1-e^{-\a})}
+\int_0^\io d\a\, \n(\a)H(\a,\ul h)$. 

Using the fact that $H_R(\a,\ul h_R)=H_R(\a,\hat\th\ul h_R)$ and that,
by Schwarz inequality,
$F_i(\a,\ul h_R)F_i(\a,\hat\th\ul h_L)\le\frac12[F_i(\a,\ul h_R)
F_i(\a,\ul h_R)+F_i(\a,\hat\th\ul h_L)$ $F_i(\a,\hat\th\ul h_L)]$, we find
\bea&&\int d\a\,\n(\a)\,H(\a,\ul h)=\label{3.15}\\
&&\quad=\int d\a\,\n(\a)\,\Big[
H_R(\a,\ul h_R)+H_R(\a,\hat\th \ul h_L)-\sum_{i=1}^2
F_i(\a, \ul h_R)\, F_i(\a,\hat\th \ul h_L)\Big]\ge\nn\\ 
&&\quad\ge\fra12\int d\a\,\n(\a)\,\Big[H\big(\a,(\ul h_L,\hat\th\ul h_L)\big)+
H\big(\a,(\hat\th\ul h_R,\ul h_R)\big)\Big]\;,\nn\eea
so that (\ref{3.12a}) is proved.

Reflecting repeatedly with respect to different bonds we end up with
\begin{equation}\int d\a\,\n(\a)\,H(\a,\ul h)\ge \fra1{M}
\sum_{i=-\fra{M}2+1}^{\fra{M}2} 
\int d\a\,\n(\a)\,H\big(\a,(h_i,h_i,\ldots,h_i,h_i)\big)\;,
\label{3.16}\end{equation}
and $E_N(M,\ul h)\ge \fra1{M}
\sum_{i=-\fra{M}2+1}^{\fra{M}2} E_N\big(M,(h_i,h_i,\ldots,h_i,h_i)\big)$,
which is an example of the {\it chessboard inequality}, see Theorem 4.1 in 
\cite{FILS78}.

\subsection{Bounds for $1<p\le2$}

We now temporarily restrict to the case $1<p\le2$. 
A key remark is that in this case there exists a $p$--dependent constant 
$K_{p}$, explicitly computable as described in Appendix \ref{A3a}, 
such that the blocks in the ground state have sizes $h_i$ all satisfying
the following apriori bound:
\begin{equation}\cases{h_i\le 12 e^J\;,\hskip1.4truecm {\rm if}\quad p=2\cr
h_i\le K_pJ^{1/(2-p)}
\qquad {\rm if}\quad 1<p<2\;,\cr}\label{3.10}\end{equation}
Eq.~(\ref{3.10}) shows in particular that, if $1<p\le 2$, the FM state is not
the ground state for any finite $J$ when $N$ is sufficiently large; 
in particular in the ground state 
the number $M$ of blocks is $M\ge 2$ (and necessarily even).
Note also that since all blocks satisfy $1\le h_i\le h_{max}$, with $h_{max}$
given by (\ref{3.10}), it must be $M\le 2N\le h_{max}M$, that is $N$ and $M$
are of the same order as $N\to \io$.

To compute the ground state energy, let us introduce the 
auxiliary partition function
\begin{equation}Q_N=\sum_{M,\ul h}^* e^{-\b E_N(M,\ul h)}\label{3.16a}
\end{equation}
where the $*$ means that we are summing only over the choices 
of $M$ and $\ul h$ compatible with the constraint $\sum_{i=-\fra{M}2+1}^{
\fra{M}2} h_i=2N$ and with the bounds (\ref{3.10}). 
We can then obtain the ground state energy per site $e_0(N)$ by taking the 
limit $-\lim_{\b\to\io}\fra1\b\log Q_N=
2Ne_0(N)$.

Using (\ref{3.16}) we get:
\begin{equation}Q_N\le 
\sum_{M,\ul h}^*\ \prod_{i=-\fra{M}2+1}^{\fra{M}2}
e^{-\b \fra1{M}E_N(M,(h_i,\ldots,h_i)\,)}\;.\label{3.17}\end{equation}
By (\ref{3.11}), we find that $\fra1{M}E_N(M,(h,\ldots,h))$ can be written 
explicitly as
\bea&&\fra1{M}E_N(M,(h,\ldots,h))=-J h+A h +2J
+\int_0^\io\fra{d\a}{\G(p)}\,\fra{\a^{p-1}e^{-\a}}{(1-e^{-\a})^2}\Bigl\{
-2\tanh\fra{\a h}2+\nn\\
&&+\fra{1}{(1-e^{-2\a N})}
\fra{(1-e^{-\a h})^2}{1+e^{-\a h}}\big(2 e^{-\a(2N-h)}+
e^{-2\a N}-e^{-\a h(M-1)}\big)\Bigr\}\;.\label{3.18}\eea
where the constant $A$ in the r.h.s. is $A=\int_0^\io 
\fra{d\a}{\G(p)}\,\a^{p-1}\fra{e^{-\a}}{(1-e^{-\a})}$. Note that  
the absolute value of the term in the last line can be bounded 
above by $K N^{-p}$, for a suitable constant $K$.  
Then, recalling the definition of $E_{per}^{(h)}(Nh)$ (see Theorem 1)
and defining $e(h)$ as 
\begin{equation}e(h)\defin\lim_{N\to\io}\frac1{2Nh}E_{per}^{(h)}(Nh)=
-J +A  +\frac{2J}h
-\frac2h\int_0^\io\fra{d\a}{\G(p)}\,\a^{p-1}\fra{e^{-\a}}{(1-e^{-\a})^2}
\tanh\fra{\a h}2\label{3.18a}\end{equation}
we find that the r.h.s. of (\ref{3.17}) and thus $Q_N$ can be bounded above by
\begin{equation}Q_N\le \sum_{M,\ul h}^*\ \prod_{i=-\fra{M}2+1}^{\fra{M}2}
e^{-\b h_i e(h_i)}e^{\b K'N^{-p}}\label{3.19}\end{equation}
Using the fact that the number of terms in the sum is less than $2^{2N}$, 
we can bound the r.h.s. of (\ref{3.19}) from above by
$2^{2N}e^{\b K'/N^{p-1}} e^{-\b 2N \tilde e_0}$, where 
$\wt e_0\=\min_{h\in\zzz^+} e(h)$, so that
we find $e_0(N)\ge \wt e_0 -K'/N^{p}$. 

It follows from the explicit expression of $e(h)$ given in (\ref{3.18a})
that $\min_{h\in\zzz^+}e(h)=e\big(h^*(J)\big)$, where the integer 
$h^*(J)$ is a piecewise constant monotone increasing function of $J$. Thus 
there is a unique integer minimizer of $e(h)$ for almost all values of $J$
which jumps at a discrete set of values of $J$. At those values of $J$ the 
minimum is attained at two consecutive integers $h^*(J)$ and $h^*(J)+1$.
As discussed below, the corresponding ground states are, asymptotically 
for $N$ large, the configurations with all blocks of the same
size, either $h^*(J)$ or $h^*(J)+1$.
If $h$ is not a minimizer, then $e(h)$ is separated by a gap from $\wt e_0$.
These properties follow from the fact that $\dpr_h e(h)=0$ has a unique 
solution for $h\in\RRR^+$ and the solution is a local minimum. 

Let us now turn to the problem of finding a lower bound for $Q_N$. 
If $N$ is divisible by $h^*$, then of course
\begin{equation}Q_N\ge e^{-\b 2N e(h^*)-\b K'N^{-p}}\label{3.21}\end{equation}
so that we find
\begin{equation}e_0(N)=\wt e_0+O\big(\fra1{N^{p}}\big)\;.\label{3.22}
\end{equation}
If $N$ is not divisible by $h^*$ the error in the r.h.s. is replaced by 
$O(N^{-1})$. This follows from the fact that we can bound $Q_N$ from below by 
restricting the sum in (\ref{3.16a}) to a configuration with all but one block
of sizes $h^*$: the last block being of size $h^*<h<2h^*$. 
The energy of such configuration is $2Ne(h^*)+O(1)$. This concludes 
the proof of Theorem 1 in the case $p\le 2$.

\subsection{The case $p>2$}

Let us now discuss the case $p>2$. It is straightforward to verify that 
in this case, if $J\ge J_0\=\int_0^\io \frac{d\a}{\G(p)}\a^{p-1}\frac{e^{-\a}}
{(1-e^{-\a})^2}$, then the minimizer of $e(h)$ is $h=+\io$. Correspondingly
we can prove that the ground state of (\ref{1.1}) is the ferromagnetic state. 
In fact let us assume by contradiction that in the ground state there is a 
block $B_h$ of finite size $h$. Then it must be true that
the energy $\D E(h)$ needed to reverse the sign of all the spins in $B_h$
is nonnegative: $\D E(h)\ge 0$. On the other side 
$\D E(h)\le -4J+2 E_1(h)$, where $E_1(h)$ is the dipole energy between 
the $+$ block $B_h$ and an external sea of $+$ spins. It is straightforward 
to check that 
\begin{equation}E_1(h)= 2\int_0^\io\frac{d\a}{\G(p)}\a^{p-1}\frac{e^{-\a}}
{(1-e^{-\a})^2}(1-e^{-\a h})<2J_0\label{3.22a}\end{equation}
and this leads to a contradiction. This proves Theorem 1 for $p>2$ and 
$J\ge J_0$. 

If on the contrary $J<J_0$, a repetition of the proof in Appendix \ref{A3a}
leads to the bound $h_i\le K_2'/(J_0-J)^{1/(p-2)}$ on the sizes $h_i$ of the 
blocks in the ground state. Then we can repeat the proof above
to get the desired result (\ref{1.2a}) and this concludes the proof 
of Theorem 1. \qed

\subsection{Uniqueness of the ground state}

In this subsection we prove Theorem 2.
If $p>2$ and $J\ge J_0$ the statement is a corollary 
of the proof above: in fact the contradiction obtained after (\ref{3.22a})
shows that in this case the ferromagnetic state is the unique ground state.

Let us then consider the cases $1<p\le 2$ or $p>2$ and $J<J_0$, in 
which we have an apriori upper bound on the sizes of the blocks.
Let us first consider the case in which 
the minimizer $h^*$ of $e(h)$ is unique and let us choose $2N=M h^*$.
Let $\ul h^0$ be a configuration in $\L_N$ for which 
the set $\III=\{i : h_i^0\neq h^*\}$ is non empty. Note that the number 
of blocks $M^0$ in $\ul h^0$ is not necessarily equal to $M$. 
Using again the chessboard estimate (see Theorem 4.1 of \cite{FILS78}), 
the energy $E_N(M^0,\ul h^0)$ 
of the configuration $\ul h^0$ can be bounded below as:
\be E_N(M^0,\ul h^0)\ge \fra1{M^0}\sum_{i=-\fra{M^0}2+1}^{\fra{M^0}2}
E_N\big(M^0,(h_i^0,\ldots, h_i^0)\big)\ge 
\sum_{i=-\fra{M^0}2+1}^{\fra{M^0}2} 
\big(h_i^0 e(h_i^0)-\fra{K}{N^p}\big)\label{3.23}\end{equation}
Using (\ref{3.22}) we find that 
\begin{equation}E_N(M^0,\ul h^0)-2N e_0(N)\ge \sum_{i\in\III} h_i^0\big(
e(h_i^0)-e(h^*)\big) +O(\fra1{N^{p-1}})\;,\label{3.24}\end{equation}
so that, since $e(h_i^0)-e(h^*)\ge \D e$, we have that for $N$ big enough
the unique ground state for $2N=M h^*$ is the state with $M$ blocks all of 
sizes $h^*$. Similarly, if $2N$ is not divisible by $h^*$, any configuration
with a sufficiently big number of ``wrong'' blocks (\ie of blocks with 
sizes $\neq h^*$), will be separated by a gap from the ground state.\\
\\
Finally, let us consider the case in which there are two minimizers 
$h^*,h^*+1$. The same proof as above goes on to show that, if $N$ is 
divisible either by $h^*$ or by $h^*+1$, then the ground state energy
is given by (\ref{3.22}) and the energy of configurations 
with blocks of ``wrong'' sizes (\ie different both from $h^*$ and $h^*+1$) 
is asymptotically larger than the energy of the ground state.
Similarly, if $N$ is not divisible either by $h^*$ or by 
$h^*+1$, the energy of configurations with a sufficiently big number of
blocks with wrong sizes will be larger than the ground state energy.\\
\\
We are left with considering the case of a configuration with blocks'
sizes all equal either to $h^*$ or to $h^*+1$. Let us denote 
by $\ss(h^*)$ the $2h^*$--periodic configuration 
with all blocks of size $h^*$ and by $\ss(h^*+1)$
the $2(h^*+1)$--periodic configuration 
with all blocks of size $h^*$. 
Can a configuration with a finite fraction of blocks of sizes $h^*$ and $h^*+1$
be a ground state? 

Let us consider a configuration $\ul h^0$ with $h_i^0\in\{h^*,h^*+1\}$ 
and the total number of blocks equal to $M$ in a volume $2N$,
with $N$ divisible by $h^*$ or $h^*+1$ and $M$ divisible by $4$
(this condition on $M$ is not restrictive: if $M$ were not divisible by $4$ 
we could consider the configuration obtained by doubling the system). Let 
\begin{equation}\III_0\defin \{i : h_i^0=h_{i+1}^0=h^*\}\;,\quad 
\III_1\defin \{i : h_i^0=h_{i+1}^0=h^*+1\}\;,\quad 
\III_2\defin \{i : h_i^0\neq h_{i+1}^0\}\;,\label{3.25}\end{equation}
where of course $h_{\fra{M}2+1}^0=h_{-\fra{M}2+1}^0$. 
Note that $|\III_0|+|\III_2|/2=M_0$, where $M_0$ is the number of blocks 
of size $h^*$, and $|\III_1|+|\III_2|/2=M_1$, 
where $M_1$ is the number of blocks of size $h^*+1$.

Using again the chessboard inequality, we find that 
\begin{equation}E_N(M,\ul h^0)\ge \fra1{M}\sum_{i=-\fra{M}2+1}^{\fra{M}2}
E_N\big(M,(h_i^0,h_{i+1}^0,h_{i+1}^0, h_i^0,\ldots,h_i^0,h_{i+1}^0,h_{i+1}^0, 
h_i^0)\big)\label{3.26}
\end{equation}
Now, except for an error of $O(M^{-(p-1)})$, the r.h.s. of (\ref{3.26}) can be 
rewritten as $|\III_0|h^*\wt e_0 +|\III_1|
(h^*+1)\wt e_0+|\III_2|(h^*+\fra12)\wt e$, where 
$\wt e$ is the (infinite volume) specific energy of 
the $(4h^*+2)$--periodic configuration 
obtained by repeating periodically the configuration $(h^*,h^*+1,h^*+1,h^*)$
over the volume. The key remark is that
$\wt e-\wt e_0=\d>0$: this is proven in Appendix \ref{A5}.
Then, using the fact that $|\III_0|+|\III_2|/2=M_0$ (with $M_0$ the number 
of blocks of size $h^*$) and that 
$|\III_1|+|\III_2|/2=M_1$ (with $M_1$ the number 
of blocks of size $h^*+1$), we get
\begin{equation}E_N(M,\ul h^0)-2N e_0(N)\ge |\III_2|(h^*+\fra12)(\wt e-\wt e_0)
\label{3.30}\end{equation} 
and the proof of Theorem 2 is concluded. \qed

\*\section{Higher dimensions}\lb{sec4}

In this section we want to prove Theorem 3. 
As already remarked after Theorem 3, for $d=2$ and $p=3$ the upper bound was 
obtained in \cite{MWRD95} by minimizing over the periodic striped 
configurations (and checking that 
asymptotically for large $J$ such an upper bound is better than the one
obtained by minimizing over the periodic checkerboard configurations).
The upper bound in dimension higher than 2 and for $d<p<d+1$ can be again 
obtained by a variational
computation, minimizing the energy over the periodic striped configurations.
We do not repeat the details here; the result is provided by the upper 
bounds in (\ref{1.3aa}) and (\ref{1.3}).

We now focus on the lower bounds in (\ref{1.3aa}) and (\ref{1.3}). 
We need to introduce some definitions; in particular 
via the basic Peierls construction we introduce the definitions of 
{\it contours} and {\it droplets}.
Given any possible configuration $\ss$ 
on $\L_N$, and in particular the ground state configuration(s), 
we define $\D$ to be the set of sites at which $\s_i=-1$, \ie
$\D=\{i\in\L\,:\,
\s_i=-1\}$. We draw around each $i\in\D$ the $2d$ sides of the unit 
$(d-1)$ dimensional cube
centered at $i$ and suppress the faces which occur twice: we obtain 
in this way a {\it closed polyhedron} $\G(\D)$ which can be thought 
as the boundary of $\D$. Each face of $\G(\D)$ separates a point 
$i\in\D$ from a point $j\not\in\D$.
Along a $d-2$ dimensional edge of $\G(\D)$ there can be either 2 or 4 faces 
meeting.
In the case of 4 faces, we deform slightly the polyhedron, ``chopping off'' 
the edge from the cubes containing a $-$ spin.
When this is done
$\G(\D)$ splits into disconnected polyhedra $\g_1,\ldots,\g_r$ which we shall 
call {\it contours}. Note that, because of the choice of periodic boundary 
conditions, all contours are closed but can possibly wind around the 
torus $\L_N$.
The definition of contours naturally induces a notion of connectedness 
for the spins in $\D$: given $i,j\in\D$ we shall say that $i$ and $j$ 
are connected iff there exists a sequence $(i=i_0,i_1,\ldots,i_n=j)$ such 
that $i_m,i_{m+1}$, $m=0,\ldots,n-1$, are nearest neighbors and none of the 
bonds $(i_m,i_{m+1})$ crosses $\G(\D)$. The maximal connected components 
$\d_i$ of $\D$ will be called {\it droplets} and the set of droplets
of $\D$ will be denoted by $\DD(\D)=\{\d_1,\ldots,\d_s\}$. 
Note that the boundaries $\G(\d_i)$ of the droplets $\d_i\in\DD(\D)$ 
are all distinct subsets of $\G(\D)$ with the property: $\cup_{i=1}^s\G(\d_i)=
\G(\D)$.

Given the definitions above, let us rewrite the energy in (\ref{1.1}) as
\begin{equation}H_N(\ss)=|\L_N|e_{FM}(J)+2J\sum_{\g\in\G(\D)}|\g|-
\sum_{\d\in\DD(\D)}
E_{dip}(\d)\;,\label{2.3}\end{equation}
where $e_{FM}(J)$ is the energy per site of the ferromagnetic configuration
$\s_i\=+1$ and $E_{dip}(\d)\defin 2\sum_{i\in\d}
\sum_{j\in\D^c}J_p(i-j)$. Let us arbitarily 
choose a number $\ell\ge 1$ (to be conveniently fixed below) and let us 
correspondingly rewrite $E_{dip}(\d)$ as
\begin{equation}E_{dip}(\d)=2\sum_{i\in\d}
\sum_{j\in\D^c}\openone(|i-j|\le \ell)J_p(i-j)+
2\sum_{i\in\d}\sum_{j\in\D^c}\openone(|i-j|> \ell)J_p(i-j)\label{2.4}
\end{equation}
Denoting the last term by $E^{>\ell}(\d)$,
we note that $\sum_{\d\in\DD(\D)}E^{>\ell}(\d)$ can be bounded above 
by $2\min\{|\D|,|\D^c|\} \Phi_p(\ell)\le |\L|\Phi_p(\ell)$, 
where $\Phi_p(\ell)=\sum_{|n|>\ell}|n|^{-p}\le const.\ell^{-(p-d)}$.
For large $\ell$ the constant is smaller than $2d|\SSS_d|/(p-d)$ 
where $|\SSS_d|$ is the volume of the $d$--dimensional unit sphere.
The first term in the r.h.s. of (\ref{2.4}) can be rewritten as
\begin{equation}E^{\le \ell}(\d)=
2\sum_{|n|\le \ell}J_p(n)\sum_{i\in\d}\sum_{j\in\D^c}
\openone(i-j=n)\le 
2\sum_{|n|\le \ell}\frac1{|n|^p}\sum_{i\in\d}\sum_{j\in\zzz^d\setminus\d}
\openone(i-j=n)\label{2.5}\end{equation}
where in the last inequality we neglected an error term vanishing in the 
thermodynamic limit.

Now, the number of ways in which $n=(n_1,\ldots,n_d)$ 
may occur as the difference $i-j$ or $j-i$ with $i\in\d$ and $j\not\in\d$
is at most $\sum_{\g\in\G(\d)}\sum_{i=1}^d|\g|_i|n_i|$, 
where $|\g|_i$ is the number of 
faces in $\g$ orthogonal to the $i$--th coordinate direction. To prove this, 
draw a path on the lattice of length $|n_1|+\cdots+|n_d|$
connecting $i$ and $j$. Such a path 
must cross a face of $\g$ and if this face is orthogonal to 
the $i$--th coordinate axis, can do so only in $|n_i|$ ways
and the claim follows. 

The conclusion is that 
\begin{equation}E^{\le \ell}(\d)
\le \sum_{\g\in\G(\d)}\sum_{|n|\le \ell}\fra1{|n|^p}
\sum_{i=1}^2 |\g|_i|n_i|=\sum_{\g\in\G(\d)}|\g|
\sum_{|n|\le \ell}\fra{|n_1|}{|n|^p}\;.\label{2.7}\end{equation}
The sum $\sum_{|n|\le \ell}|n_1|/|n|^p$ can be bounded above by
$\int_{1\le|x|\le\ell}d^dx|x_1|/|x|^{p}+const.$ where
the constant is smaller than $2pd|\SSS_d|2^{p-d}/(p-d)$ and  
$|\SSS_d|$ is the volume of the $d$--dimensional unit sphere.
Then (\ref{2.7}) implies 
$E^{\le \ell}(\d)\le 2\Psi_p(\ell)+2pd|\SSS_d|2^{p-d}/(p-d)$, where 
$\Psi_p(\ell)=|\SSS_{d-1}|\log\ell$ if $p=d+1$ and 
$\Psi_p(\ell)=|\SSS_{d-1}|(\ell^{d+1-p}-1)/(d+1-p)$ if $d<p<d+1$. 

Putting these bounds back into (\ref{2.3}), we find:
\begin{equation}E_0(N)\ge |\L_N|e_{FM}+2\left(
J-\frac{pd|\SSS_d|2^{p-d}}{p-d}-
\Psi_p(\ell)\right)\sum_{\g\in\G(\D)}|\g|-
\fra{2d|\SSS_d|}{p-d}\frac{|\L_N|}{\ell^{p-d}}\;,\label{2.9}\end{equation}
where $\D$ is the set of $-$ sites in the (unknown) ground state configuration.
Choosing $\ell$ in such a way that $\Psi_p(\ell)=J-pd|\SSS_d|2^{p-d}/(p-d)$, 
we find the lower bounds in (\ref{1.3aa}) and (\ref{1.3}). 
This concludes the proof of Theorem 3. \qed

\\

As a side remark, let us note that a discussion similar to the one above 
implies that if $p>d+1$ and $J$ is large enough no droplet can 
appear in the ground state, \ie the ground state is ferromagnetic. 
In fact, let us assume by contradiction that 
the ground state is not $\s_i\=-1$ and that 
there is at least one droplet $\d$ in the ground state. Then the energy
needed to reverse all the spins in $\d$ must be positive: 
$-2J\sum_{\g\in\G(\d)}|\g|+
E_{dip}(\d)\ge0$. Proceeding as above in the proof of (\ref{2.7}) 
and using that 
$p>d+1$, we get $E_{dip}(\d)\le \sum_{\g\in\G(\d)}|\g|
\sum_{|n|\ge 1}\fra{|n_1|}{|n|^p}\le const. \sum_{\g\in\G(\d)}|\g|$; hence 
for $J$ big enough we get a contradiction and this proves that
the ground state is ferromagnetic.\\

The contour method implemented in the proof of Theorem 3 also allows one 
to get some informations
about the geometry of contours and droplets in the ground state for $d<p\le 
d+1$. In fact if $\DD(\D)$ is the set of droplets in the ground state and 
we consider a configuration $\ss'$ on $\{\pm1\}^\L$
with $\D'$ s.t. $\DD(\D')=\DD(\D) \setminus\{\d\}$, for some $\d$, 
then it must be $H_N(\ss')-E_0(N)\ge 0$. 
With the definitions introduced above we have 
\bea H_N(\ss')-E_0(N)&&\le -2J\sum_{\g\in\G(\d)}|\g|+E_{dip}(\d)\le\nn\\ 
&&\le2\left(-J+\frac{pd|\SSS_d|2^{p-d}}{p-d}+\Psi_p(\ell)\right)
\sum_{\g\in\G(\d)}|\g|+
\fra{2d|\SSS_d|}{p-d}\frac{|\d|}{\ell^{p-d}}\;.\label{2.12}\eea
Imposing that the r.h.s. of (\ref{2.12}) is positive and choosing 
$\Psi_p(\ell)+\frac{pd|\sss_d|2^{p-d}}{p-d}-J=-1$,
we find $\sum_{\g\in\G(\d)}|\g|\le const.|\d|e^{-J/|\sss_{d-1}|}$ 
if $p=d+1$ and $\sum_{\g\in\G(\d)}|\g|\le const.|\d|J^{-\frac{p-d}{d+1-p}}$ 
if $d<p<d+1$. If $\d$ does not wind up $\L_N$,
using the isoperimetric inequality $|\d|\le(\sum_{\g\in\G(\d)}|\g|/2d)^d$
we also find that in the ground state $\sum_{\g\in\G(\d)}|\g|\ge const.
e^{J/|\sss_{d-1}|}$ if $p=d+1$ or $\sum_{\g\in\G(\d)}|\g|\ge 
const.J^{\frac{p-d}{d+1-p}}$ if $d<p\le d+1$.\\

Finally we mention that exploiting the same energy arguments above, one can 
prove that the ground state of Hamiltonian (\ref{1.1}) with $\L_N$ an 
$N^{d-1}\times w$ cylinder with periodic boundary conditions is the 
ferromagnetic ground state, {\it provided} $w\le K_3' e^{J/|\sss_{d-1}|}$, 
if $p=d+1$, or $w\le K_p' J^{\frac{p-d}{d+1-p}}$ if $d<p\le d+1$, 
for a suitable constant $K_p'$. This result 
strongly suggests that for $d<p\le d+1$ the droplets in the ground state
are quasi--$(d-1)$ dimensional structures of width $O(e^{J/|\sss_{d-1}|})$ 
or $O(J^{\frac{p-d}{d+1-p}})$, reminiscent of the conjectured stripes. 
If we could prove that the droplets 
in the ground state must necessarily be stripes, then we could exploit 
the methods in Section.\ref{sec3} to prove that the ground state is indeed 
realized by a periodic striped configuration.

\*\section{Concluding remarks}\lb{sec5}

\subsection{One dimension.}

{\it Ground state.}
In 1D the characterization of the infinite system ground state is complete:
the ground state is periodic and there is a sequence of 
transitions from an antiferromagnetic state of period two to $2h$--periodic 
states of blocks of sizes $h$, $h>1$, with alternating sign, the size $h$ 
changing (increasing discontinuously) when the ferromagnetic interaction 
strength $J$ is increased.

Our proof is based on a reflection positivity argument which relies heavily
on the details of the model Hamiltonian, \eg on the fact that 
the long--range repulsion is reflection positive. 
It is an interesting open problem to establish more general 
conditions the long range repulsive interaction should satisfy in order to 
guarantee existence of a periodic modulated ground state. 
Note that the problem of determining the ground state of a spin system with 
positive and convex potential was solved by Hubbard and by
Pokrovsky and Uimin (even in presence of 
a magnetic field) \cite{Hu78,PU78}: this means that for $J=0$ the assumption of
convexity of the potential is enough to determine the ground state of 
(\ref{1.1}). The proof in \cite{Hu78,PU78}
was generalized to an ``almost'' convex case by Jedrzejewski and Miekisz
\cite{JM001,JM002}. However this proof requires a small ferromagnetic
interaction $1\le J\le 1+2^{-p}$. It is an open problem 
to generalize our analysis (in the presence of a large 
ferromagnetic coupling) to the case of a long range convex 
interaction. 

It would also be interesting to establish what would be the 
effect of adding a magnetic field to the model Hamiltonian. It is expected 
that in the presence of a large short range ferromagnetic coupling 
and of a positive magnetic field
$B$ of increasing strength the ground state is still periodic with the 
$+$ blocks larger than the $-$'s, at least if $|B|$ is not too large 
\cite{GD82}. However for large magnetic 
field it could be the case that small variations of the magnetic field could
induce an infinite sequence of transitions between 
periodic states characterized by different rational values of the 
magnetization (Devil's staircase): this is in fact what happens for $J=0$ 
as a function of the magnetic field \cite{BB82}. According
to \cite{BB82} the antiferromagnetic state would remain 
the ground state for $|B|<B_c$ for $J=0$. For larger values of $B$ 
there is a sequence of transitions from the antiferromagnetic state of period 
two to (complicated) periodic states with all possible rational values of the 
magnetization.

{\it Positive temperature.}
Another interesting open problem in one dimension consists in establishing 
properties of the infinite volume Gibbs measures at positive 
temperatures. Note that for a {\it ferromagnetic} long range interaction 
with decay $1/r^p$, $1<p\le 2$, it is known \cite{D69,FS82} that 
there is a phase transition for the inverse temperature
$\b$ large enough. In the case
of model (\ref{1.1}), on the contrary, it is natural 
to expect a unique Gibbs measure for any finite $\b$, even for $1<p\le 2$.
It is in fact known that if $J=0$ in (\ref{1.1}), then for $d=1$ and 
$p>1$ there is a unique limit Gibbs state at all values of the inverse 
temperature $\b$ \cite{FVV82,BS83,K99}. The proof in \cite{K99} does not 
extend to the case $J>0$. It is known however that for $J>0$ all Gibbs states
are translation invariant \cite{FVV82}. Similarly, using the argument in 
\cite{BLP79}, one can show that the Gibbs states obtained as limits of 
finite volume Gibbs measures with boundary conditions 
$\t^k\ul\s_{per}(h^*(J))$, where $\ul\s_{per}(h^*(J))$ is a periodic 
ground state configuration with blocks of size $h^*(J)$ and $\t$ the 
translation operator, are all equivalent among each other, for any 
$k=1,\ldots,h^*(J)$. If, on the contrary, $0<p\le 1$ and, say, $J=0$,
it has been conjectured \cite{Kh}
that model (\ref{1.1}), modified as explained after (\ref{1.1}),
admits at least two different Gibbs states obtained as limits of 
finite volume Gibbs measures with boundary conditions $\cdots +-+- \cdots$
and $\cdots -+-+ \cdots$ (which are presumably well defined because the 
Hamiltonian is thermodynamically stable, see Remark (1) after Theorem 2).

\subsection{Higher dimensions.}

{\it Ground state.}
In two or more dimensions we have shown that 
the specific ground state energy agrees 
asymptotically for large $J$ with the best variational ground state 
known so far, which is a periodic striped configuration \cite{MWRD95}. 

As already mentioned in the introduction, the ground states are believed,
on the basis of variational computations and MonteCarlo simulations,
to consist of a sequence of ``stripes'' of size $h$ 
(growing when the ferromagnetic interaction strength $J$ is increased) 
and alternating signs.

If $J$ is sufficiently large, the addition of a magnetic field is believed  
to lead to a thickening of the stripes with spins parallel to the external 
field. At very large fields a transition from a striped to a ``bubble'' phase 
is expected \cite{GD82}: the bubble phase consisting of large ``cilindrical'' 
droplets of spins parallel to the external field arranged in a periodic 
fashion and surrounded by a ``sea'' of spins of opposite sign. 

{\it Positive temperature.}
The non trivial structure of the set of ground
states is believed to have a counterpart at positive temperature: in 
particular in the absence of magnetic field it is
believed that at low temperature there are different 
pure Gibbs states describing 
striped states with two possible orientations (horizontal and vertical).
The striped states are expected to ``melt'' at a positive critical temperature
\cite{SS99} with the ``stripe melting'' described in terms of an effective 
Landau--Ginzburg free energy functional \cite{M02}. 

It goes without saying that it would be of great interest to substantiate 
these pictures by rigorous proofs. Since 
most standard methods (cluster expansion, correlation inequalities) seem to 
fail in giving any useful information for systems of spins with long range 
antiferromagnetic interactions, new ideas are needed to understand 
some of the aforementioned problems.

\appendix
\section{A priori bounds on the size of the blocks}\lb{A3a}
\setcounter{equation}{0}
\renewcommand{\theequation}{\ref{A3a}.\arabic{equation}}

In this appendix we prove (\ref{3.10}). Given a block $B_{h_i}$ of size $h_i$,
let us assume that $h_i=(2K+1)h$, 
for two integers $K$ and $h$ to be chosen conveniently
(it will be clear from the proof below that this assumption is not 
restrictive). To be definite let us assume
that the block $B_{h_i}$ of size $h_i$ we are considering is 
a $+$ block. Since $h_i=(2K+1)h$, we can think of 
$B_{h_i}$ as a sequence of three
contiguous blocks of sizes $Kh,h$ and $Kh$ respectively, to be called 
$B_h^1,B_h^2,B_h^3$. 
Let us now compute the energy needed to flip the block $B_h^2$
and let us call it $\D E'(h)$. If we are in the ground state, then of course 
$\D E'(h)\ge 0$. Also, we have $\D E'(h)\le 4J-2E_1(h)$,
where $E_1(h)$ is the dipole interaction energy 
between a $+$ block $B_h^2$ of size $h$ and a configuration of spins in 
$\ZZZ\setminus B_h^2$ such that the spins in $B_h^1,B_h^3$ are all $+$
and the spins in $\ZZZ\setminus B_{h_i}$ are all $-$.
$E_1(h)$ is readily computed as
\bea E_1(h)&&=2\sum_{i=1}^h\Big[
\sum_{j=h+1}^{h+Kh}-\sum_{j\ge h+Kh+1}
\Big]\frac1{(j-i)^p}=\nn\\
&&=\sum_{i=1}^h\Big[\sum_{j=h+1}^{h+Kh}-\sum_{j\ge h+Kh+1}
\Big]\int_0^\io \frac{d\a}{\G(p)}\a^{p-1}e^{-\a(j-i)}=\nn\\
&&=2\int_0^\io\fra{d\a}{\G(p)}\,\a^{p-1} \fra{e^{-\a}}
{(1-e^{-\a})^2}(1-e^{-\a h})
(1-2e^{-\a Kh})\label{A3a.4}\eea
so that, using the condition $\D E'(h)\ge 0$, we find that in the ground state
\begin{equation}\int_0^\io\fra{d\a}{\G(p)}\,
\a^{p-1} \fra{e^{-\a}}{(1-e^{-\a})^2}
(1-e^{-\a h})
(1-2e^{-\a Kh})\le J\label{A3a.5}\end{equation}
Using now the fact that $\a e^{-\a/2}\le 1-e^{-\a}\le\a$ for $\a\ge 0$, 
the l.h.s. of (\ref{A3a.5}) can be bounded below by:
\begin{equation}\fra1{\G(p)}\int_0^\io \fra{d\a}{\a}\fra1{\a^{2-p}}
e^{-\a}(1-e^{-\a h})
(1-2e^{-\a (Kh-1)})\label{A3a.7}\end{equation}
Now, if $p=2$ (\ref{A3a.7}) can be computed to give
$\log(h+1)/(1+1/K)^2$. Then, choosing 
$K=1$ and plugging these bounds back into (\ref{A3a.5}), we find 
that, if $p=2$, $\log h\le J+2\log 2$ and $h_i=3h\le 12 e^{J}$.

If $1<p<2$, (\ref{A3a.7}) can be further bounded below as
\bea&&
\fra1{\G(p)}\int_0^\io \fra{d\a}{\a}\fra1{\a^{2-p}}\Bigl\{
e^{-\a h}(1-e^{-\a h})^2- e^{-\a Kh}(1-e^{-\a h})\Bigr\}\ge\nn\\ 
&&\ge \fra{h^{2-p}}{\G(p)}
\int_0^\io \fra{d\a}{\a}\fra1{\a^{2-p}}\Bigl\{
e^{-\a}(1-e^{-\a})^2-\a e^{-\a K}\Bigr\}\=\big(A'-\fra{B'}{K^{p-1}}\big)h^{2-p}
\label{A3a.8}\eea
Choosing $K$ in such a way that $B'/K^{p-1}\le A'/2$ and 
plugging the bounds into 
(\ref{A3a.5}), we find that $A' h^{2-p}\le 2J$, so that $h_i=(2K+1)h\le 
(2K+1)(2J/A')^{1/(2-p)}$.

\section{The integral representation}\lb{A4}
\setcounter{equation}{0}
\renewcommand{\theequation}{\ref{A4}.\arabic{equation}}

In this Appendix we want to give the details of the computation leading to 
(\ref{3.11}). 
First of all let us note that the first two terms in the r.h.s. of
(\ref{3.11}) come from the short range FM interaction. In order to show
that the computation of the dipole interaction energy leads to the 
remaining terms appearing in the r.h.s. of (\ref{3.11}), 
let us first note that 
the infinite volume dipole interaction energy $E_{12}$ 
between two blocks of signs $\e_1$ and $\e_2$, $\e_\a=\pm1$,
of sizes $h_1$ and $h_2$ and 
separated by a string of $d$ spins, can be computed as follows:
\bea E_{12}
&&=\e_1\e_2\sum_{i=1}^{h_1}\sum_{j=1}^{h_2}\fra1{(j-i+h_1+d)^p}
=\e_1\e_2\int_0^\io\fra{d\a}{\G(p)}\,\a^{p-1}e^{-\a(j-i+h_1+d)}=\nn\\
&&=\e_1\e_2\int_0^\io\fra{d\a}{\G(p)}\,\a^{p-1}
\fra{e^{-\a}}{(1-e^{-\a})^2}(1-e^{-\a h_1})e^{-\a d}
(1-e^{-\a h_2})\label{A4.1}\eea
Now, in order to compute the finite volume 
dipole interaction energy of two different blocks $h_i$ and $h_j$, $j>i$,
we can simply apply the definition and (\ref{A4.1}) to find:
\bea&&\e_i\e_j\sum_{n\in \zzz}\sum_{l=1}^{h_i}\sum_{m=1}^{h_j}
\fra1{|m-l+h_i+\cdots+h_{j-1}+2nN|^p}=\label{A4.2}\\
&&=\e_i\e_j\int_0^\io\fra{d\a}{\G(p)}\,\a^{p-1}
\fra{e^{-\a}}{(1-e^{-\a})^2}(1-e^{-\a h_i})
(1-e^{-\a h_j})\Big[\prod_{i<k<j}e^{-\a h_k}\Big]
\sum_{n\ge 0}e^{-2\a nN}+\nn\\
&&+\e_i\e_j\int_0^\io\fra{d\a}{\G(p)}\,\a^{p-1}
\fra{e^{-\a}}{(1-e^{-\a})^2}(1-e^{-\a h_i})
(1-e^{-\a h_j})e^{-\a(2N-h_i-\cdots-h_j)}\sum_{n\ge 0}e^{-2\a nN}
\;,\nn\eea
which leads to the term in the third line of 
(\ref{3.11}) (we used that $\e_i\e_j=(-1)^{j-i}$ and $\sum_{i=1}^M h_i=2N$).
Similarly, the finite volume dipole self--interaction of a block $h_i$ 
can be computed as 
\bea&&\sum_{k=1}^{h_i-1}\fra{h_i-k}{k^p}+2\sum_{n\ge 1}
\sum_{i,j=1}^{h_i}\fra1{(j-i+2nN)^p}=\int_0^\io\fra{d\a}{\G(p)}\,\a^{p-1}
\fra{e^{-\a}}{(1-e^{-\a})^2}\cdot\lb{A4.3}\\
&&\cdot\Big[h_i(1-e^{-\a})
-(1-e^{-\a h_i})+2 e^{\a h_i} (1-e^{-\a h_i})^2
\fra{e^{-2\a N}}{1-e^{-2\a N}}\Big]\nn\eea
Summing (\ref{A4.3}) over $i=1,\ldots,M$, we get the last term in the first 
line of (\ref{3.11}) and the term in the second line.

\section{Computation of the 1D energy gap}\lb{A5}
\setcounter{equation}{0}
\renewcommand{\theequation}{\ref{A5}.\arabic{equation}}

In this Appendix we want to compute the specific energy $\wt e$
of the $(4h^*+2)$--periodic configuration $(h^*,h^*+1,h^*+1,h^*,\ldots)$
in the case in which $h^*$ and $h^*+1$ are both minimizers of $e(h)$.
In particular we will show that $\wt e-\wt e_0=\d>0$.
Using the general expression (\ref{3.11}), after some algebra we find that 
\bea \wt e&&=-J+A+\fra{4J}{2h^*+1}-\fra2{2h^*+1}\int_0^\io d\a\,\n(\a)
\Biggl\{\fra{2(1-e^{-\a h^*})(1-e^{-\a (h^*+1)})}
{1-e^{-\a(4h^*+2)}}+\nn\\
&&+\fra{e^{-2\a h^*}(1-e^{-\a (h^*+1)})^2+e^{-2\a(h^*+1)}(1-e^{-\a h^*})^2}
{1-e^{-\a(4h^*+2)}}\Biggr\}\label{A5.1}\eea
where $\n(\a)\=(\G(p))^{-1}\a^{p-1}
e^{-\a}(1-e^{-\a})^{-2}$.
We want to prove that this expression is strictly larger than $\wt e_0$.
Note that, since both $h^*$ and $h^*+1$ are minimizers, we have
\begin{equation}\fra{J}{h^*}-\fra1{h^*}\int_0^\io d\a\,
\n(\a)\tanh\fra{\a h^*}2=
\fra{J}{h^*+1}-\fra1{h^*+1}\int_0^\io d\a\,\n(\a)\tanh\fra{\a (h^*+1)}2
\label{A5.2}\end{equation}
implying that 
\begin{equation}J=\int_0^\io d\a\,\n(\a)\Big[(h^*+1)\tanh\fra{\a h^*}2-h^*
\tanh\fra{\a (h^*+1)}2\Big]\label{A5.3}\end{equation}
and 
\begin{equation}
\wt e_0=2\int_0^\io d\a\,
\n(\a)\Big(\tanh\fra{\a h^*}2-\tanh\fra{\a (h^*+1)}2\Big)
\label{A5.4}\end{equation}
Using (\ref{A5.3}) and (\ref{A5.4}) we find
\bea&&\wt e-\wt e_0=2\int_0^\io d\a\,\n(\a) \Bigg[
\tanh\fra{\a h^*}2+\tanh\fra{\a (h^*+1)}2-\lb{A5.5}\\
&&-\fra{2(1-e^{-\a h^*})(1-e^{-\a (h^*+1)})+
e^{-2\a h^*}(1-e^{-\a (h^*+1)})^2+e^{-2\a(h^*+1)}(1-e^{-\a h^*})^2}
{1-e^{-\a(4h^*+2)}}\Bigg]\nn\eea
A bit more of algebra shows that (\ref{A5.5}) can be rewritten as
\begin{equation}\wt e-\wt e_0=2\int_0^\io d\a\,\n(\a) 
(e^{-\a h^*}-e^{-\a(h^*+1)})^2
(1-e^{-\a h^*})(1-e^{-\a (h^*+1)})\label{A5.6}\end{equation}
and this concludes the proof.

\acknowledgments 

We would like to thank G. Gallavotti, T. Kuna, K. Khanin and A.C.D. van Enter
for useful discussions and comments. 
The work of JLL was supported by NSF Grant DMR-044-2066 and by AFOSR
Grant AF-FA 9550-04-4-22910
and was completed during a visit to the IAS. The work of AG and EHL 
was partially supported by U.S. National Science Foundation
grant PHY 01 39984, which is gratefully acknowledged.


\end{document}